\documentclass[sigconf, natbib=false]{acmart}

\AtBeginDocument{%
  }

\usepackage{amsmath,amsfonts}
\usepackage{algorithmic}
\usepackage{graphicx}
\usepackage{textcomp}
\usepackage{subcaption}
\usepackage{svg}
\usepackage{booktabs}
\usepackage{tablefootnote}
\usepackage{tabularx}
\usepackage{booktabs}  
\usepackage{makecell}
\usepackage[table]{xcolor}

\RequirePackage[
  datamodel=acmdatamodel,
  style=acmnumeric,
  ]{biblatex}

\begin{document}

\title{Bridging Simulation and Silicon: A Study of RISC-V Hardware and FireSim Simulation}

\author{Atanu Barai$^1$, Kamalavasan Kamalakkannan$^1$, Patrick Diehl$^1$, Maxim Moraru$^1$, Jered Dominguez-Trujillo$^1$, Howard Pritchard$^1$, Nandakishore Santhi$^1$, Farzad Fatollahi-Fard$^2$, Galen Shipman$^1$}

\affiliation{
    \institution{
    $^1$ Los Alamos National Laboratory, Los Alamos, NM \country{USA}\\
    $^2$ Lawrence Berkeley National Laboratory, Berkeley, CA \country{USA}\\
    \{abarai, kamalavasan, diehlpk, moraru, jereddt, howardp, nsanthi, gshipman\}@lanl.gov, ffard@lbl.gov
    }
}

\begin{abstract}
RISC-V ISA-based processors have recently emerged as both powerful and energy-efficient computing platforms. The release of the MILK-V Pioneer marked a significant milestone as the first desktop-grade RISC-V system. With increasing engagement from both academia and industry, such platforms exhibit strong potential for adoption in high-performance computing (HPC) environments.

The open-source, FPGA-accelerated FireSim framework has emerged as a flexible and scalable tool for architectural exploration, enabling simulation of various system configurations using RISC-V cores. Despite its capabilities, there remains a lack of systematic evaluation regarding the feasibility and performance prediction accuracy of FireSim when compared to physical hardware.

In this study, we address this gap by modeling a commercially available single-board computer and a desktop-grade RISC-V CPU within FireSim. To ensure fidelity between simulation and real hardware, we first measure the performance of a series of benchmarks to compare runtime behavior under single-core and four-core configurations. Based on the closest matching simulation parameters, we subsequently evaluate performance using a representative mini-application and the LAMMPS molecular dynamics code.

Our findings indicate that while FireSim provides valuable insights into architectural performance trends, discrepancies remain between simulated and measured runtimes. These deviations stem from both inherent limitations of the simulation environment and the restricted availability of detailed performance specifications from CPU manufacturers, which hinder precise configuration matching.
\end{abstract}


\renewcommand{\shortauthors}{A.~Barai, K.~Kamalakkannan, P.~Diehl, M.~Moraru, J.~Dominguez-Trujillo, H.~Pritchard, N.~Santhi, F.~Fatollahi-Fard, G.~Shipman}

\maketitle
\vspace{-0.6em}
\section{Introduction}
\vspace{-0.4em}
With the rise of open hardware ecosystems, the RISC-V instruction set architecture has emerged as a compelling alternative to proprietary ones, offering extensibility, transparency, and a growing community of academic and industrial support, and stimulating significant interest within the high-performance computing (HPC) community. As the low-cost RISC-V development boards and advanced simulation platforms become more available, researchers now have a rich collection of tools to explore architectural trade-offs in high-performance scientific computing.

In order to maximize performance and efficiency gains, HPC researchers and system designers increasingly leverage simulation-based methodologies, such as the FPGA accelerated cycle-accurate FireSim~\cite{FiresimKarandikar} platform, to rapidly prototype and evaluate architectural innovations prior to tape-out. However, the fidelity and modeling ability of such simulation frameworks for full system simulations, and the extent to which they accurately predict performance of real-world scientific applications on actual RISC-V hardware, remains an open question. On the other hand, physical boards such as RISC-V Banana Pi and MILK-V often provide limited flexibility in terms of observability and configurability.

This paper seeks to addresses this gap in research by comparing and contrasting the performance and behavior of multiple simulated and physical RISC-V based CPUs. Specifically, it investigates the ability of the FireSim simulation framework to model two RISC-V CPUs broadly available in the market using a variety of microbenchmarks and scientific HPC benchmarks. These benchmarks were run a variety of simulated RISC-V CPUs as well as two commercial RISC-V CPU products, Banana Pi and MILK-V. By quantifying differences in performance, scalability, and computational efficiency, it provides insights into the strengths and limitations of simulation-based approaches for RISC-V based HPC hardware development and assessment.

We aim to empower HPC researchers and system architects by analyzing and clarifying how well the current generation FireSim simulation tool models and reflects real hardware performance. Thus, we enable informed decisions on future HPC system design and the potential adoption of RISC-V architectures in scientific computing contexts by quantifying their opportunities and limitations. In summary, we make the following contributions in this paper:
\begin{enumerate}
    \item We evaluate the capability of FireSim to closely model and to predict the performance on single core and multi-core systems.
    \item We evaluate the performance of HPC representative workloads on simulated and actual systems and analyze the correlations in performance.
\vspace{-5pt}
\end{enumerate}

\section{Related work}
Berger-Vergiat \emph{et al.}~\cite{wl-char-on-firesim-sandia} focused on porting and evaluating a few HPC software stacks on different single-core RISC-V simulation configurations. Specifically, they simulated Rocket, BOOM, and CVA6 core-based processors on FireSim. In contrast, we focus on evaluating FireSim itself by closely matching the FireSim simulation configurations with commercially available multicore in-order and out-of-order hardware platforms and comparing the performance for scientific applications.


Multi-core RISC-V systems were also simulated using gem5~\cite{ta2018simulating}. Memory-Bound Kernels were executed on RISC-V CPUs and compared with Arm and x86 CPUs in~\cite{10.1007/978-3-031-41673-6_5}. The real-time performance of RISC-V is studied in~\cite{yoo2024real}. Adaptive simulation on an open-source RISC-V platform was evaluated using virtual prototypes in~\cite{herdt2021adaptive}. Deep learning convolutions based on GEMM were evaluated on ARM and RISC-V architectures~\cite{MARTINEZ2024103186}. Diehl et.\ al ran the real world application Octo-Tiger on the MILK-V nodes and got comparable performance with Riken's supercomputer\ Fugaku~\cite{diehl2023evaluating,10820750}. In addition, they benchmarked the vectorization on the Banana Pi single-board computer.

Dorflinger \emph{et al.}~\cite{dorflinger-risc-v-survey} performed detailed evaluation of single-core Rocket, BOOM, CVA6, and SHAKTI C-Class RISC-V implementations on both FPGA and ASIC platform. They evaluated several metrics including processing performance, area and resource utilization, power consumption, and efficiency. They found that Rocket cores outperform the others for all evaluation criteria, except for processing performance on the ASIC platform.  Höller \emph{et al.}~\cite{holler} presented a structured overview and hardware-focused evaluation of open-source 32-bit RISC-V cores for FPGA implementation. They identify key selection criteria, such as a free ISA, open-source licensing, standard bus interfaces, and compiler support for sixteen candidate cores implementing three representative designs on a Xilinx Artix-7 FPGA, with results showing trade-offs between resource usage, clock frequency, and performance. Although these works analyzed different RISC-V cores for area and power, none attempted to evaluate the performance alignment of workloads between RISC-V processors and simulation. In another study, Brown \emph{et al.}~\cite{MilkV2023} evaluated the 64-core Sophon SG2042 RISC-V CPU performance against an x86 counterpart and found that the x86 processor outperforms SG2042 by 4-8 times for multi-threaded workloads.

In this work, we study the performance of the workloads on a simulated RISC-V model for various architectural parameters and provide insights on how architectural parameters influence specific workload's performance. In contrast to prior works, we leverage the FireSim framework to model commercially available RISC-V processors, aligning the simulated model’s architectural parameters with the target hardware platforms available in the market. We evaluate the configurability of FireSim’s hardware models for hardware and performance matching and highlight their limitations.

\section{Experimental Approach and Setup}

In this study, we focus on single core and multi-core performance of applications on FireSim while closely matching the simulated system's configuration with real hardware to compare the performance results of software workloads across simulation and hardware. To evaluate, we compare the FireSim simulated platform with the Banana Pi BPI-F3 SpacemiT K1 hardware platform. We use the in-order Rocket cores~\cite{rocket} to configure FireSim simulations as similar to Banana Pi as possible. We utilize the out-of-order soft BOOM~\cite{sonicboom} cores to model the out-of-order SOPHON SG2042-based MILK-V processor. To tune the Rocket and BOOM cores to closely match the performance and microarchitectural features of the Banana Pi and MILK-V hardware, respectively, we utilize a set of 40 microbenchmarks~\cite{microbench} which target individual micro-architectural features. Once we have tuned our simulated cores, we choose three scientific applications - UME, NAS Parallel Benchmark (NPB), and Lammps to represent compute-, memory-, and network-bound applications.

Next, we describe the hardware and software setup for our experiments in detail. First, we describe the hardware simulation platform, FireSim, and the corresponding commercially available hardware platforms we attempt to model. Then we present the benchmarks and workloads used to evaluate the hardware and simulate FireSim models.

\vspace{-0.7em}
\subsection{Platforms}

\subsubsection{FireSim}

FireSim~\cite{FiresimKarandikar} is an open-source, cycle-accurate framework that maps hardware modules onto FPGAs and links the targets through a scalable distributed network, enabling multi-node full-system simulations at MHz-level speeds. These large-scale simulations enable early-stage performance assessment and prediction for a wide range of system architectures. FireSim distribution comes with few open-source CPU cores such as the in-order Rocket~\cite{rocket} core, and out-of-order, superscalar BOOM~\cite{sonicboom} core which are developed as part of Berkeley Architecture Research (BAR) project. These cores support running full Linux systems with 64-bit floating point capabilities. Both cores are written in Chisel, a hardware construction language embedded in Scala, which facilitates parameterization and design reuse. While Rocket is well-suited for control-oriented and real-time applications, BOOM targets compute-intensive workloads, making the two cores complementary within the RISC-V ecosystem.

In our experiments, we instantiate multi-core configurations of these CPU cores without enabling vector units and make use of MPI to run parallel applications. We use the Chipyard~\cite{chipyard} tool to instantiate the processors keeping cache size, memory bandwidth and other characteristics similar to the real hardware we compare against. To accomplish this, we used the Berkeley eXtensible Environment (BXE)~\cite{bxe-firesim-wiki}, a Chipyard/FireSim environment and 22x AMD EPYC 7282 16-Core Processor/Xilinx Alveo U250 FPGA~\cite{amd-alveo-u250-a64g} cluster hosted at Lawrence Berkeley National Laboratory. BXE provides the necessary tools for building hardware designs with Chipyard and the associated software to run in the simulation. The finalized design is simulated in FireSim on the aforementioned FPGAs.

\vspace{-0.5em}
\subsubsection{Banana Pi}
In 2024, the Banana Pi F3 became the first single-board computer to feature the 8-core SpacemiT K1 processor, which implements 256-bit vector processing fully compliant with the RISC-V Vector Extension Version 1.0 (V1.0) specification. The eight cores are organized in two clusters where each cluster has four cores with 32K L1-Cache per core, 512K L2-Cache. It also has a real-time CPU system responsible for handling different communication peripherals such as I2C, UART, PWM, SPI \emph{etc}. It has just 4GB of memory. It runs Bianbu Linux 1.0.13, a board support package provided by SpacemiT which is based on Debian Linux running kernel 6.1.15. The Banana Pis are hosted at Louisiana State University. When running the workloads on Banana Pi, we use only one cluster with 4-core by binding the processes to those cores only. 

\vspace{-0.5em}
\subsubsection{MILK-V}
The MILK-V pioneer is the first desktop-grade RISC-V computer available. The Pioneer Box is a ready desktop computer equipped with 128GB DDR4, 1 $\times$ 1TB PCIe 3.0 SSD, 1 $\times$ AMD R5 230 Graphic Card, and 1 $\times$ Intel X540-T2 Network Card (10Gbps). The board contains one RISC-V SOPHON SG2042 CPU with 64 cores (2GHz) and L1 Cache of 64KB, L2 Cache of 1MB/Cluster with 4 cores per Cluster, and L3 cache of 64MB. We ran Fedora 38 with Kernel version 6.1.55. The system is hosted at Louisiana State University and fully integrated in the research cluster \texttt{rostam}. 

\subsection{Software Workloads}

\subsubsection{MicroBench}

\begin{table*}[!htb]
    \centering
    \caption{MicroBench kernels, categories, and descriptions}
    \vspace{-1em}
    \label{tab:microbench}
    \rowcolors{2}{gray!25}{white}
    \begin{tabular}{l|l|l||l|l|l}
    \toprule
    Name & Category & Description & Name & Category & Description \\
    \midrule
    Cca & Control Flow & Completely biased branch & EM1 & Execution & Int -- Length 1 dependency chain \\
    Cce & Control Flow & Alternating branches & EM5 & Execution & Int -- Length 5 dependency chain  \\
    CCh & Control Flow & Random control flow & MC & Cache & Conflict misses \\
    CCh\_st & Control Flow & Impossible to predict control + stores & MCS & Cache & Conflict misses with stores \\
    CCl & Control Flow & Impossible control w/ large Basic Blocks & MD & Cache & Cache resident linked list traversal \\
    CCm & Control Flow & Heavily biased branches  & MI & Cache & Independent access, cache resident \\
    CF1 & Control Flow & Inlining test for functions w/ loops & MIM & Cache & Independent access, no conflicts  \\
    CRd & Control Flow & Recursive control flow -- 1000 Deep & MIM2 & Cache & Independent access -- 2 coalescing ops \\
    CRf & Control Flow & Recursive control flow - Fibonacci & MIP & Cache & Instruction cache misses \\
    CRm & Control Flow & Merge sort & ML2 & Cache & L2 linked-list \\
    CS1 & Control Flow & Switch -- Different each time & ML2\_BW\_ld & Cache & L2 linked-list -- B/W limited (lds) \\
    CS3 & Control Flow & Switch -- Different every third time  & ML2\_BW\_ldst & Cache & L2 linked-list -- B/W limited (ld/sts) \\
    DP1d & Data & Data parallel loop - Double arithmetic & ML2\_BW\_st & Cache & L2 linked-list -- B/W limited (sts) \\
    DP1f & Data & Data parallel loop -- Float arithmetic & ML2\_st & Cache & L2 linked-list (sts) \\
    DPT & Data & Data parallel loop -- Sin() & STL2 & Cache & Repeatedly store, L2 resident \\
    DPTd & Data & Data parallel loop -- Double sin() & STL2b & Cache & Occasional stores, L2 resident\\
    DPcvt & Data & Data parallel loop -- Float to Double & STc & Cache & Repeated consecutive L1 store \\
    ED1 & Execution & Int -- Length 1 dependency chain  & M\_Dyn & Cache & Load store w/ dynamic dependencies\\
    EF & Execution & FP -- 8 Independent instructions  & MM & Memory & Non-cache resident linked-list\\
    EI & Execution & Int -- 8 Independent computations & MM\_st & Memory & Non-cache resident linked-list (sts) \\
    \bottomrule
    \end{tabular}
\end{table*}

The MicroBench suite, inspired by~\cite{microbench}, provides 40 lightweight microbenchmarks specifically targeted at evaluating microarchitectural features of processor cores and memory subsystems (Table \ref{tab:microbench}). These benchmarks encompass tests for control flow behavior (\emph{e.g.}, branch prediction), arithmetic operations (compute capability), and diverse memory access patterns (\emph{e.g.}, cache bank conflicts, memory access latency, and overall memory subsystem bandwidth). We employ this suite to pinpoint microarchitectural bottlenecks in our simulation model, as recommended in~\cite{nowatzki}, and subsequently tune the model parameters to achieve close alignment with observed hardware performance. 39 of the 40 benchmarks were used in our evaluation, since CRm resulted in a segfault on all simulated and real hardware.

\subsubsection{The NAS Parallel Benchmarks}
The NAS Parallel Benchmarks (NPB)\cite{bailey1991nas} are a set of standardized tests designed to evaluate the performance of massively parallel supercomputers. It was developed by NASA Ames Research Center. These benchmarks are based on computational fluid dynamics (CFD) applications and are intended to represent typical computation and communication patterns in large-scale scientific computing. The NPB suite includes both synthetic and application-based benchmarks, such as IS (Integer Sort), EP (Embarrassingly Parallel), CG (Conjugate Gradient), and MG (Multi-Grid). These represent different computational and communication patterns, including random memory access, irregular memory access and communication, all-to-all communication, and parallel and distributed computing domains. Each application is designed to stress different aspects of a parallel system's architecture, including processor speed, memory hierarchy, and interconnect bandwidth and latency. Available in both MPI and OpenMP versions, the NPBs facilitate performance comparison across diverse high-performance computing platforms. Their widespread use has established them as a critical tool in the architectural evaluation and optimization of modern parallel systems.

We executed the NAS Parallel Benchmarks, version 3.4.2. Table \ref{tab:nas:list} lists the benchmarks and their respective classes. The experiments were conducted on both the physical hardware platform and simulation models for Class A, using 1 and 4 MPI ranks. Class A was selected because it can be run on actual hardware in roughly ten seconds, while its simulation takes on the order of few hours. The longer simulation time arises from the FPGA-based models operating at 60 MHz for the Rocket core (approximately 25× slower than a representative 1.6 GHz system) and 15 MHz for the BOOM core (around 135× slower than a representative 2.0 GHz system). This slowdown, relative to actual hardware, is further intensified by FireSim’s token-based simulation models for DRAM and LLC, which deliberately stall cores and memory to maintain the target execution frequency.

\begin{table}[htb]
    \centering
     \caption{NPB apps used in the experiments}
     \vspace{-1em}
    \label{tab:nas:list}
    \begin{tabular}{l|l|l}\toprule
     Benchmark & Characteristics~\cite{MilkV2023} & Class \\
     \midrule
    \rowcolor{lightgray} CG & Memory Latency & A \\
    EP & Compute & A \\
     \rowcolor{lightgray} IS & Memory Latency, BW  & A \\
    MG & Memory Latency, BW & A \\\bottomrule
    \end{tabular}
\end{table}

\subsubsection{UME}
The UME (Unstructured Mesh Explorations)~\cite{ume} proxy application, developed at Los Alamos National Laboratory, models salient features of a much larger code. Specifically, UME represents only a few specific, but commonly used algorithms, taking particular care to reproduce the data management and memory access patterns found in the original application. This open-source, C\texttt{++}20-based proxy application facilitates the exploration and optimization of unstructured mesh representations and their resultant memory access patterns, which are crucial to the performance of many numerical methods in multiphysics codes at LANL. 



In contrast to the implicit connectivity of a structured mesh, unstructured mesh representations require feature connectivity to be described and stored explicitly. When we consider the existence of mesh features such as subzonal volumes, the number of possible connectivity maps rapidly grows out of control. To tame this growth, connectivity hierarchies are employed. These hierarchies allow us to explicitly represent only a subset of all possible conductivities, but they lead to multi-level indirection. Due to the indirection through connectivity maps, typical loop structures in UME lead to very high integer operation counts, very high load/store ratios, and low floating-point intensity. 





In our experiment we chose the input set as $32^3$ 3-dimensional mesh (32,768 total zones). The number of other entities scales with the number of zones, with about 8 corners per zone, about 12 edges per zone, about 8 points per zone,  about 6 faces per zone, \emph{etc}.


\subsubsection{Lammps}
The Large-scale Atomic/Molecular Massively Parallel Simulator (Lammps)~\cite{thompson2022lammps} is a parallel molecular dynamics simulation package developed by Sandia National Laboratories, specifically engineered to run efficiently on high-performance computing platforms. It supports a diverse array of interatomic potentials and force fields, including Lennard-Jones, EAM, Tersoff, ReaxFF, and coarse-grained models, making it suitable for a wide range of applications such as solid-state materials, metals, polymers, biological macromolecules, and soft condensed matter.

Lammps is specifically engineered for high-performance computing (HPC) environments, making it highly effective for large-scale molecular dynamics simulations. Its parallelization strategy is based on spatial domain decomposition combined with message-passing interface (MPI) communication, allowing it to scale efficiently across thousands of processors on distributed-memory systems. LAMMPS also supports multi-threading via OpenMP and accelerates performance further with GPU computing through CUDA, HIP, or the Kokkos performance portability library. These features enable LAMMPS to fully exploit modern heterogeneous HPC architectures. Additionally, its modular design allows selective compilation of performance-critical packages, minimizing computational overhead. Efficient load balancing, optimized neighbor list construction, and support for hybrid MPI/OpenMP/GPU execution models contribute to Lammps’s strong scalability and versatility on supercomputers. These capabilities make it a powerful tool for simulating large systems, performing long time-scale simulations, and integrating seamlessly into coupled multi-physics workflows in demanding research and industrial applications. In our experiment, we ran the \emph{Lennard Jones} and \emph{Chain} benchmarks from Lammps with standard inputs enabling MPI.

\subsubsection{Software Packages Used}
Table~\ref{tab:compilers} shows the used compilers and application versions. Note that the compiler versions are different for the simulation and on the hardware. This is due to the fact that FireSim comes with Ubuntu-20.04 and supports GCC-9.4.0. Upgrading GCC on FireSim to version 13.2 requires building it from source code which is time-consuming.

\section{Firesim Modeling of Banana Pi and MILK-V}
\label{sec:modeling}

\begin{table}[htbp]
    \centering
    \caption{Compiler and application versions.}
    \vspace{-1em}
    \label{tab:compilers}
    \begin{tabular}{lc||ll}\toprule
    Hardware & gcc/openmpi & Application & Version  \\\midrule
    \rowcolor{lightgray}  MILK-V  & 13.2.1/5.0.3 & MicroBench/NAS & \textit{9b9cdba}/3.4.2  \\
     Banana Pi  & 13.2.0/3.1 & UME  & \textit{7b235b5}\\
   \rowcolor{lightgray}   Fire-sim & 9.4.0 & LAMMPS & \textit{0a6b13ff0b} \\\bottomrule
    \end{tabular}
\end{table}

\begin{table*}[htbp]
    \centering
    \caption{FireSim Models}
    \vspace{-1em}
    \begin{tabularx}{\textwidth}{l|l|l|l|l|l|l|l}
        \toprule
        FireSim Model & Clock Rate & Front End & RoB & LSQ & L1D/I Cache & L2 Banks & System bus \\
        \midrule
        Rocket 1 & 1.6 GHz & Fetch:2, Decode:1 & N/A & N/A & Sets:64, Ways:8 & 1 & 64-bit\\
        Rocket 2 & 1.6 GHz & Fetch:2, Decode:1 & N/A & N/A & Sets:64, Ways:8 & 4 & 128-bit\\
        Small BOOM & 2.0 GHz & Fetch:4, Decode:1 & RoB:32 & Load:8, Store:8 & Sets:64, Ways:4 & 4 & 128-bit\\
        Medium BOOM & 2.0 GHz & Fetch:4, Decode:2 & RoB:64 & Load:16, Store:16 & Sets:64, Ways:4 & 4 & 128-bit\\
        Large BOOM & 2.0 GHz & Fetch:8, Decode:3 & RoB:96 & Load:24, Store:24 & Sets:64, Ways:8 & 4 & 128-bit\\
        \bottomrule
    \end{tabularx}
    \label{tab:firesim_models}
    \vspace{-1em}
\end{table*}

To accurately model a commercially available System-on-Chip (SoC), detailed micro-architectural information is essential. Additionally, the simulation framework must support customizable models of architectural components. However, processors from industrial vendors often lack disclosure of key micro-architectural details—such as a core's reorder buffer size, the number of physical registers, and similar internal structures. Instead, vendors typically provide only high-level architectural specifications, including core count, cache hierarchy, and external memory configuration.

Given this limitation, we configured the high-level architectural parameters in our simulation to align with those of the actual hardware. To refine the model further, we conducted empirical experiments using microbenchmarks to identify performance differences. Based on these insights, we tuned the micro-architectural parameters to more closely replicate the behavior of the target processor. As a result, we developed simulated models that closely mirror the performance characteristics of the Banana Pi and MILK-V hardware platforms. Architectural specification of the hardware and their corresponding FireSim models are provided in Table~\ref{tab:arch_spec}. 

We are modeling a single cluster of the Banana Pi's SpacemiT K1 chip in FireSim, by leveraging the \emph{huge rocket core} configuration available in the rocket-chip~\cite{rocket} generator. We chose the rocket core configuration as both the rocket core and SpacemiT K1 core are in-order processors. Moreover, \emph{huge rocket core} configuration matches the L1 D+I cache sizes of the Banana Pi's SpacemiT K1 cores. We also kept the default 512 KiB L2 cache for the rocket tile such that it represent a single Banana Pi cluster. The primary known difference between our simulation model and the actual Banana Pi hardware lies in the external memory configuration, pipeline stages and issue width. While the Banana Pi employs dual 32-bit LPDDR4 memory at 2666 Mbps, our simulation uses a 2000 Mbps DDR3 FR-FCFS quad-rank model, as FireSim currently only provides direct support for DDR3 memory models. The cores in the Banana Pi’s SpacemiT K1 chip feature an 8-stage pipeline, whereas our simulation model, which uses the Huge Rocket core, implements only a 5-stage pipeline. Additionally, the SpacemiT K1 cores are designed for dual-issue execution, while the Rocket core is a single-issue, in-order processor.

 Our \texttt{Rocket1} configuration described in Table~\ref{tab:firesim_models} is equvalent to Huge Rocket core in chipyard repository. In the \texttt{Rocket2} configuration, the number of cache banks was increased from one (in \texttt{Rocket1}) to four. The \texttt{Banana Pi Sim Model} configuration additionally employed a wider bus (128 bits instead of 64). To mimic the dual issue execute in simulation, we doubled the modeled frequency to 3.2 GHz, and refer to this configuration as the \texttt{Fast Banana Pi Sim Model}. 
 

To model the out-of-order execution cores found in the MILK-V Pioneer, we evaluated three core configurations from the BOOM repository~\cite{riscv-boom-github}: Small, Medium, and Large. We executed both microbenchmarks and NAS parallel benchmarks across these configurations. Based on empirical results, the Large BOOM core most closely approximates the performance characteristics of the MILK-V Pioneer’s cores for compute-intensive workloads. We adopted the Large BOOM configuration from the repository and modified it to reflect the MILK-V's cache hierarchy: increasing both the L1 instruction and data caches to 64 KiB, and adding a 1 MiB L2 cache. To model the 64 MiB last-level cache (LLC) of the MILK-V Pioneer, we used four 16 MiB LLCs, each connected to one of FireSim's four memory channels. It is important to note that FireSim's LLC model is simplified—it behaves like an SRAM and does not account for detailed cache system latencies such as tag access delay or data retrieval latency. Due to resource limitation of available FPGAs, we modeled only one cluster with 4 cores. In contrast, MILK-V pioneer has 64 cores. 

\renewcommand{\arraystretch}{1.1}
\begin{table*}[t]
\centering
\caption{Architecture Specification of Tuned FireSim Models}
\vspace{-1em}
\begin{tabularx}{\textwidth}{p{2.4cm} | p{2.8cm} p{3.5cm} | p{2.8cm} p{5cm}}
\toprule
\multicolumn{1}{c} {} & \multicolumn{2}{|c}{\textbf{Banana Pi}} & \multicolumn{2}{|c}{\textbf{MILK-V Pioneer}} \\
\cmidrule(lr){2-3} \cmidrule(lr){4-5}
\textbf{Parameter} & \textbf{Hardware} & \textbf{(Fast) Simulation Model} & \textbf{Hardware} & \textbf{Simulation Model} \\
\midrule
 SoC & SpacemiT K1 & Rocket Tile & SOPHON SG2042 4 & Boom Tile \\
Core Count & 8 & 4 & 64 & 4 \\
 Core Frequency & 1.6 GHz & (3.2 GHz) 1.6 GHz & 2.0 GHz & 2.0 GHz \\
Core Execution & In-order & In-order & Out of Order & Out of Order \\
 TLB size & N/A & \makecell[lt]{L1 D, I - 32 entry \\ (fully associative)}  & N/A & \makecell[lt]{L1 D, I - 32 entry (fully associative) \\ L2 -1024 entry (direct mapped)} \\
FrontEnd & N/A  & \makecell[lt]{2-wide fetch, 1-wide decode\\
BTB,BHT,RAS-branch  \\ predictors} & N/A & \makecell[lt]{8-wide fetch, 3-wide decode \\ TAGE-L branch predictor  \\ 24-entry fetch buffer, \\ 16 outstanding branches} \\
Execute & Dual Issue & \makecell[lt]{Single Issue} & N/A & \makecell[lt]{96-entry rob, 16-entry 1-issue \\ memory queue, 32-entry \\ 3-issue integer queue, 24-entry \\ 1-issue fp queue} \\
LSQ & N/A & None & N/A & 24-entry load/store queue \\ 
\cmidrule(lr){2-5}
& \multicolumn{4}{c} {Memory sub system} \\
\cmidrule(lr){2-5}
 L1 D,I Cache size & 32KiB & 32KiB & 64 KiB & 64 KiB \\
L1 D sets/ways & N/A & 64/8 & N/A & 128/8 \\
 L1 I sets/ways & N/A & 64/8 & N/A & 128/8 \\
L2 Cache (Cluster) & 512KiB (4 cores) & 512 KiB (4 Cores) & 1 MiB (4 Cores) & 1 MiB (4 Cores) \\
 L2 sets/ways & N/A & 1024/8 & N/A & 1024/8 \\
System Bus & N/A & 128 bit & N/A & 128 bit \\
 LLC & None & None & 64 MiB & 64 MiB \\ 
External Memory & \makecell[lt]{Dual 32-bit \\ LPDDR4 2666Mbps} &  \makecell[lt] {DDR3 2000 Mbps \\ FR-FCFS quad-rank}  & \makecell[lt]{4 channel \\ DDR4 3200 Mbps} & \makecell[lt] {4 channel DDR3 2000 Mbps \\  FR-FCFS quad-rank} \\
\bottomrule
\end{tabularx}

\label{tab:arch_spec}
\end{table*}
\renewcommand{\arraystretch}{1}

\vspace{-0.7em}
\section{Results}
In this section, we use relative speedup as the primary metric to assess how closely the simulation matches the performance of the target hardware. A relative speedup of 1.2, for example, indicates that the simulation runs 20\% faster than the real hardware. Our goal is a relative speedup of 1.0, which would represent an exact performance match.

\vspace{-0.8em}
\subsection{MicroBench}
\begin{figure*}[tbh!]
  \centerline{\includegraphics[width=\textwidth]{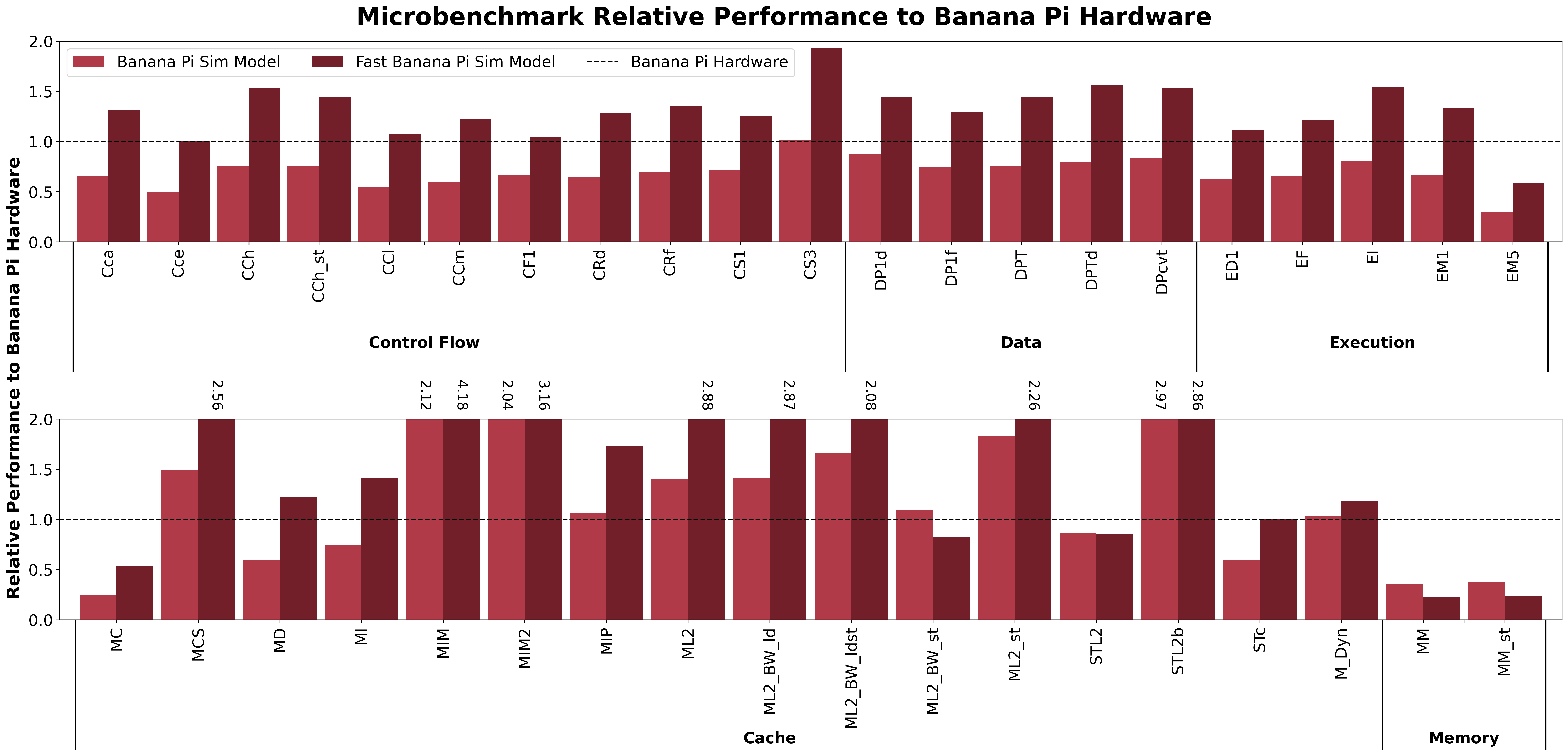}}
  \caption{Microbenchmark performance on FireSim tuned Rocket core to match the Banana Pi hardware, labeled as the \texttt{Banana Pi Sim Model} (See Table \ref{tab:arch_spec}), normalized by the Banana Pi hardware performance results. \texttt{Fast Banana Pi Sim Model} results have also been included to study the effects of increasing the clock rate by a factor of $2$ up to $3.2$GHz.}
  \label{fig:microbench-bpi}
\end{figure*}

As mentioned in Section \ref{sec:modeling}, we first chose the available Huge Rocket Core configuration as the in-order core that would best approximate the in-order cores of the Banana Pi. The main differences between the chosen configuration and the Banana Pi are the memory type and speed, and the available specs for both the chosen model and the Banana Pi hardware are in Table~\ref{tab:arch_spec}. 

As demonstrated in Figure~\ref{fig:microbench-bpi}, these differences in memory type (DDR-3 for the modeled \texttt{Banana Pi Simulation Model} vs. DDR-4 for the \texttt{Banana Pi Hardware}) and unknown parameters in memory subsystem (Cache configuration and System bus parameters) and dual issue of Banana Pi result in the simulated model only achieving $35$-$37$\% of the performance of the baseline hardware in the two microbenchmarks (MM, MM\_st) that stress the DRAM bandwidth by performing non-cache resident linked-list traversals. 
The control flow, data parallel, and execution microbenchmarks underachieve compared to the Banana Pi hardware pretty uniformly, indicating that the Banana Pi hardware may have larger and more optimized micro-architectural features. We believe this could be the dual issue execute of  the Banana Pi hardware compared to single issue execute in rocket cores used for the \texttt{Banana Pi Sim Model}. Relative performance for the \texttt{Fast Banana Pi Sim Model} indicates better matching performance for control flow, data and execution categories in MicroBench. Cache category performance is improved further as expected as cache hierarchy performance also improves when we doubles the frequency, while the memory benchmarks, which are bottlenecked by DRAM performance, experienced further drops due to the higher clock frequency becoming unbalanced with the available DRAM bandwidth, resulting in longer queues and increased latencies.

\begin{figure*}[tbh!]
  \centerline{\includegraphics[width=\textwidth]{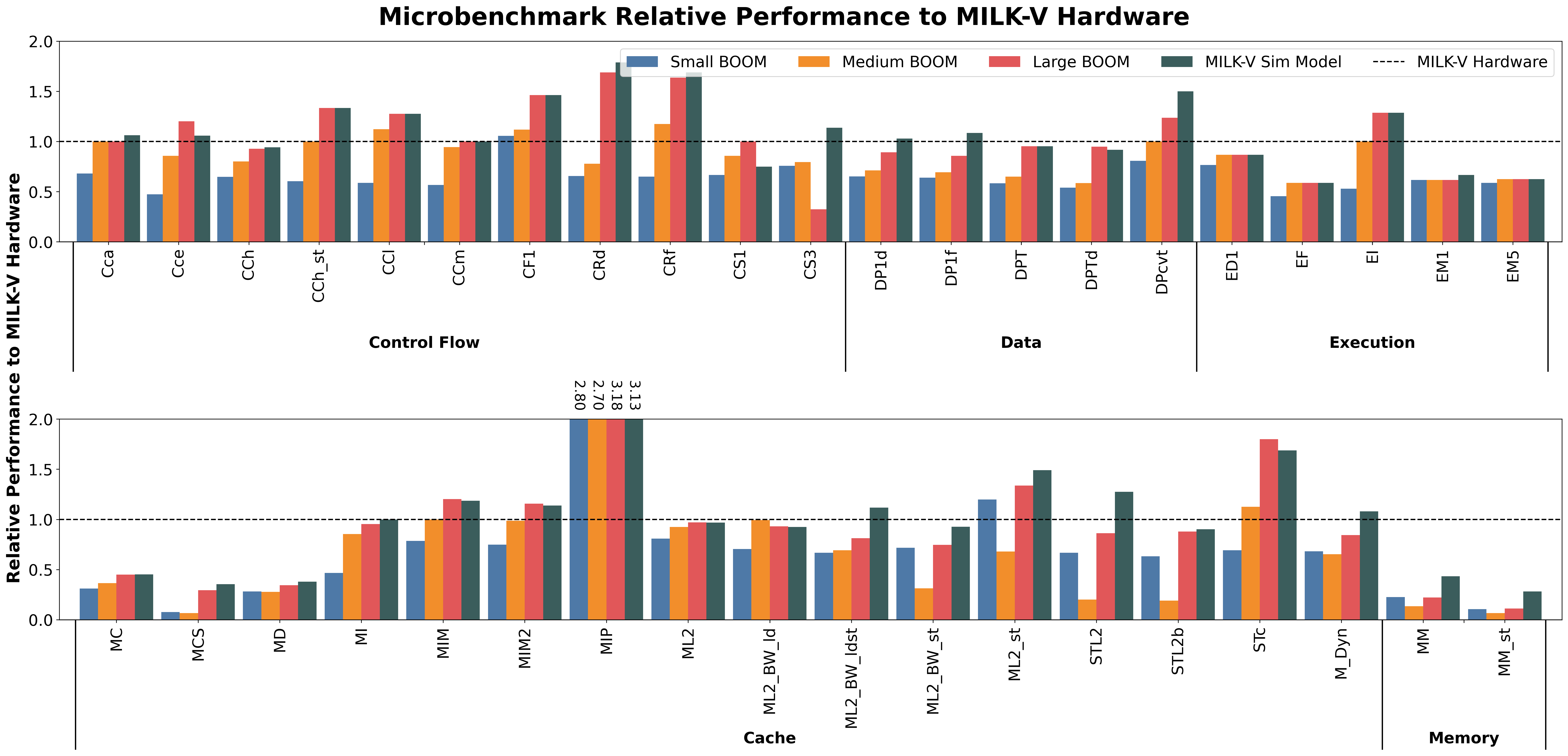}}
  \caption{Microbenchmark performance on FireSim of Small, Medium, and Large BOOM cores and a tuned Large BOOM core to match the MILK-V hardware (\texttt{MILK-V Sim Model}, See Table \ref{tab:arch_spec}), normalized by the MILK-V hardware performance.}
  \label{fig:microbench-milk-v}
\end{figure*}

To determine which out-of-order core would best model the MILK-V processor, we evaluated the performance of the small, medium, and large BOOM cores (Table \ref{tab:firesim_models}) on the microbenchmarks with FireSim. As seen in Figure~\ref{fig:microbench-milk-v}, the large BOOM core aligned best with the MILK-V, and it was determined that modifying the \texttt{Large BOOM} core to match the cache capacities of the MILK-V hardware would provide the most accurate model given the limited public information of the MILK-V micro-architecture and pipeline. This modified \texttt{Large BOOM} configuration is referred to as the \texttt{MILK-V Simulation Model} in the remainder of this paper.

Similar to the Banana Pi results, Figure~\ref{fig:microbench-milk-v} indicates that the different memory types between the \texttt{MILK-V Simulation Model} on FireSim and the MILK-V hardware greatly affected the performance of the memory microbenchmarks, with the simulated model only achieving 28\%-43\% of the baseline performance compared to the MILK-V hardware. Furthermore, the cache microbenchmarks achieve mixed results. Microbenchmarks which exercise the ability to performantly handle conflict misses perform worse on the simulation model, while all other cache benchmarks, except for the instruction cache misses microbenchmark (MIP), perform similarly to the hardware. The MIP benchmark substantially outperforms the hardware when run on FireSim for all BOOM variants. 

The control flow and data parallel microbenchmarks perform similarly between the \texttt{MILK-V Simulation Model} and the \texttt{MILK-V Hardware}, achieving relative performances from $0.75$-$1.78$, where the recursive control flow benchmarks, CRd and CRf, achieve the best performance on the FireSim \texttt{MILK-V Simulation Model}. The 4 control flow and data parallel microbenchmarks which the FireSim \texttt{MILK-V Simulation Model} under performs the \texttt{MILK-V Hardware} are CS1, CCh, DPTd, and DPcvt microbenchmarks. Finally, the execution microbenchmarks for the FireSim \texttt{MILK-V Simulation Model} indicate general under-performance. These benchmarks largely test the processor's ability to handle dependency chains, and the only microbenchmark which doesn't, EI, performs comparably with the hardware, indicating that the \texttt{MILK-V Hardware} likely contains more fetch and decode units that were modeled.

\subsection{NAS}
\subsubsection{Rocket}
We evaluated the four different configurations described  in Tables~\ref{tab:arch_spec},\ref{tab:firesim_models} on the Nas Parallel Benchmarks.

\begin{figure}
  \centering
  \subcaptionbox{single core\label{fig:npb_procs_1}}
                {\includegraphics[width=0.5\textwidth]{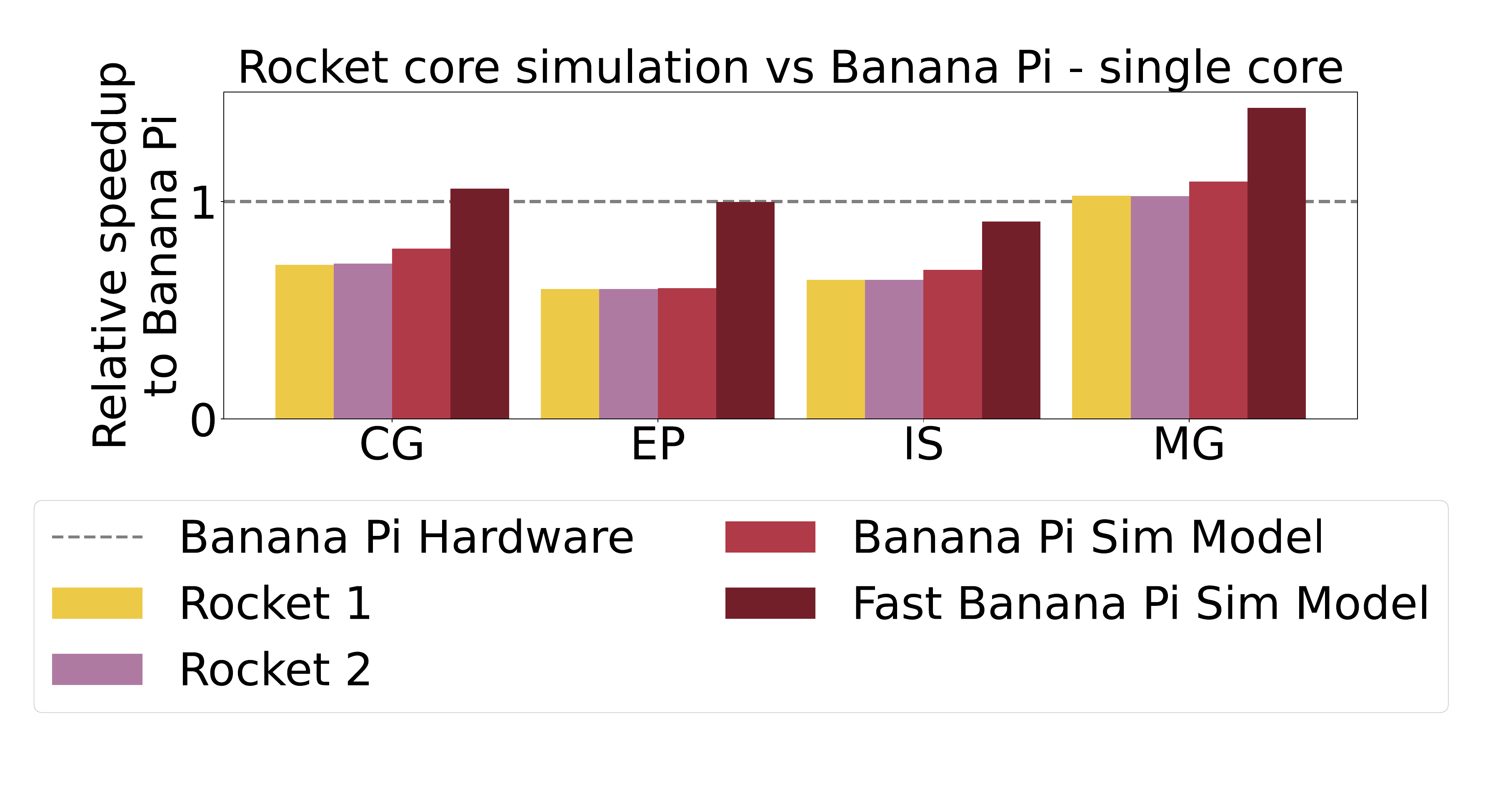}}%
  \par\smallskip
  \subcaptionbox{four cores\label{fig:npb_procs_4}}
                {\includegraphics[width=0.5\textwidth]{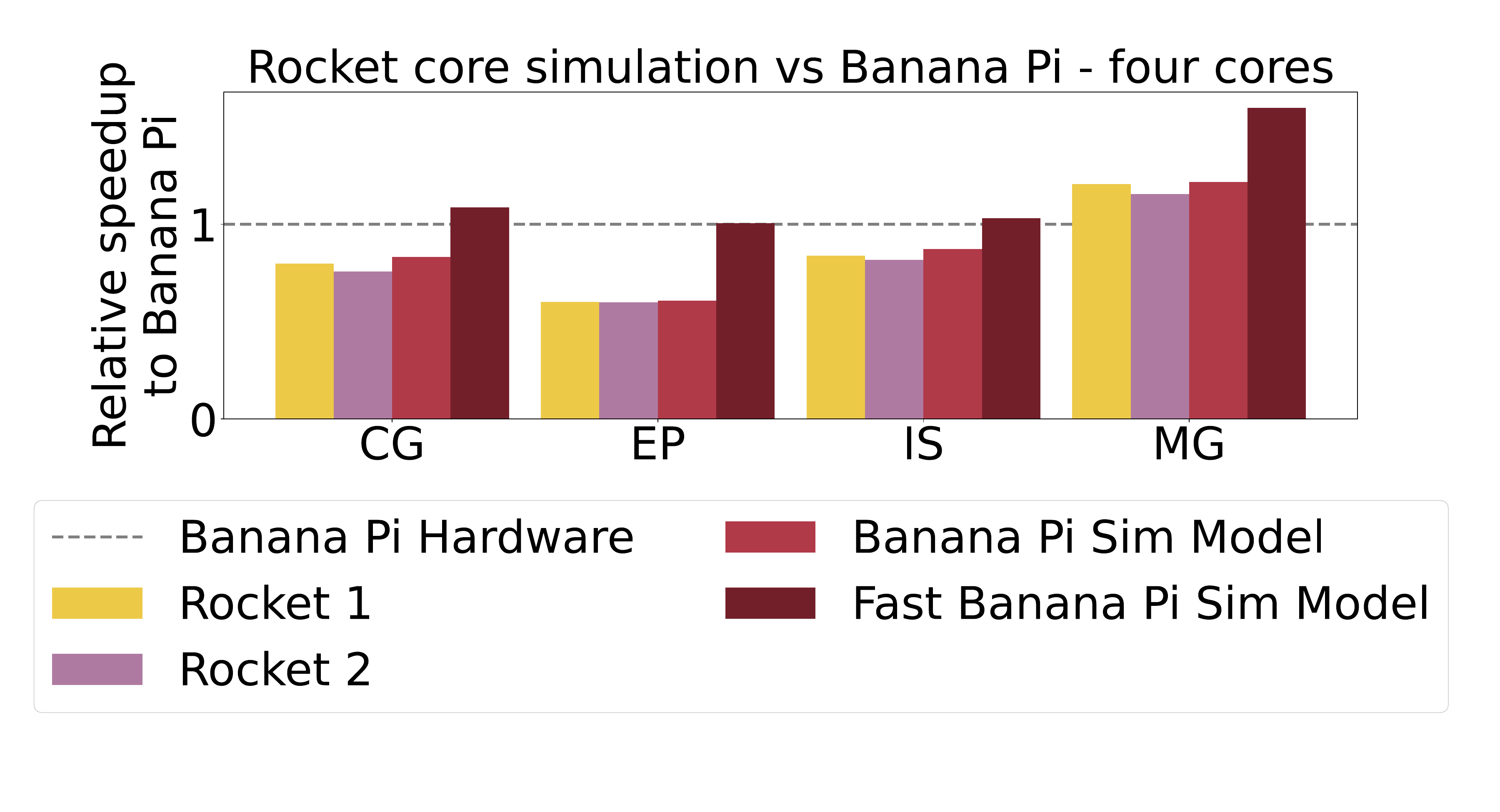}}

  \caption{Relative speedup of NAS Parallel Benchmarks on FireSim (Rocket cores) compared to Banana Pi hardware (target value = 1.0; lower deviation indicates closer performance match).}
  \label{fig:rocket_npb}
\end{figure}

Figure~\ref{fig:rocket_npb} shows the results for single and four core runs. We observed no significant performance difference between the \texttt{Rocket1} and \texttt{Rocket2} configurations. In contrast, the \texttt{Banana Pi Sim Model} configuration delivered improvements for the CG and IS benchmarks. 

The runtimes for benchmarks sensitive to cache latency and memory bandwidth (IS, CG, and MG) were reasonably close to those on the actual hardware. This is consistent with the MicroBench results, where the simulated hardware matched or outperformed the real system in cache-related tests. Performance for some benchmarks remains lower on Rocket despite its better cache-level performance. We attribute this to differences in compute capabilities and DRAM characteristics.

We observed a higher runtime for EP on the simulated hardware compared to Banana Pi, again matching the MicroBench findings. Specifically, the Rocket cores performed worse in the control flow, data, and execution categories measured by MicroBench. 

Finally, the \texttt{Fast Banana Pi Sim Model} achieved performance most closely matching the actual hardware. Again, we believe this is because doubling the frequency produces behavior similar to the dual-issue execution in the Banana Pi. However, doubling the frequency also reduces cache latency and increases throughput. Consequently, benchmarks such as MG, which are highly sensitive to cache latency and memory bandwidth, run faster on the simulated hardware, leading to performance results that no longer match those of the real device.



\subsubsection{Boom}
For the BOOM cores, we began our analysis by running several initial configurations available in the riscv-boom repository~\cite{riscv-boom-github}: small, medium, and large. The primary differences between these configurations lies in their micro-architectural resources, such as decode, issue, and fetch widths; the number of registers; the sizes of the load and store queues; the reorder buffer size; branch-handling capabilities; and other related parameters. These initial experiments were intended to identify the configuration whose performance most closely matches the MILK-V hardware.

\begin{figure}[ht]
  \centering
  \subcaptionbox{BOOM configurations from \cite{riscv-boom-github}%
    \label{results:boom_npb:init}}{%
    \includegraphics[width=0.5\textwidth]{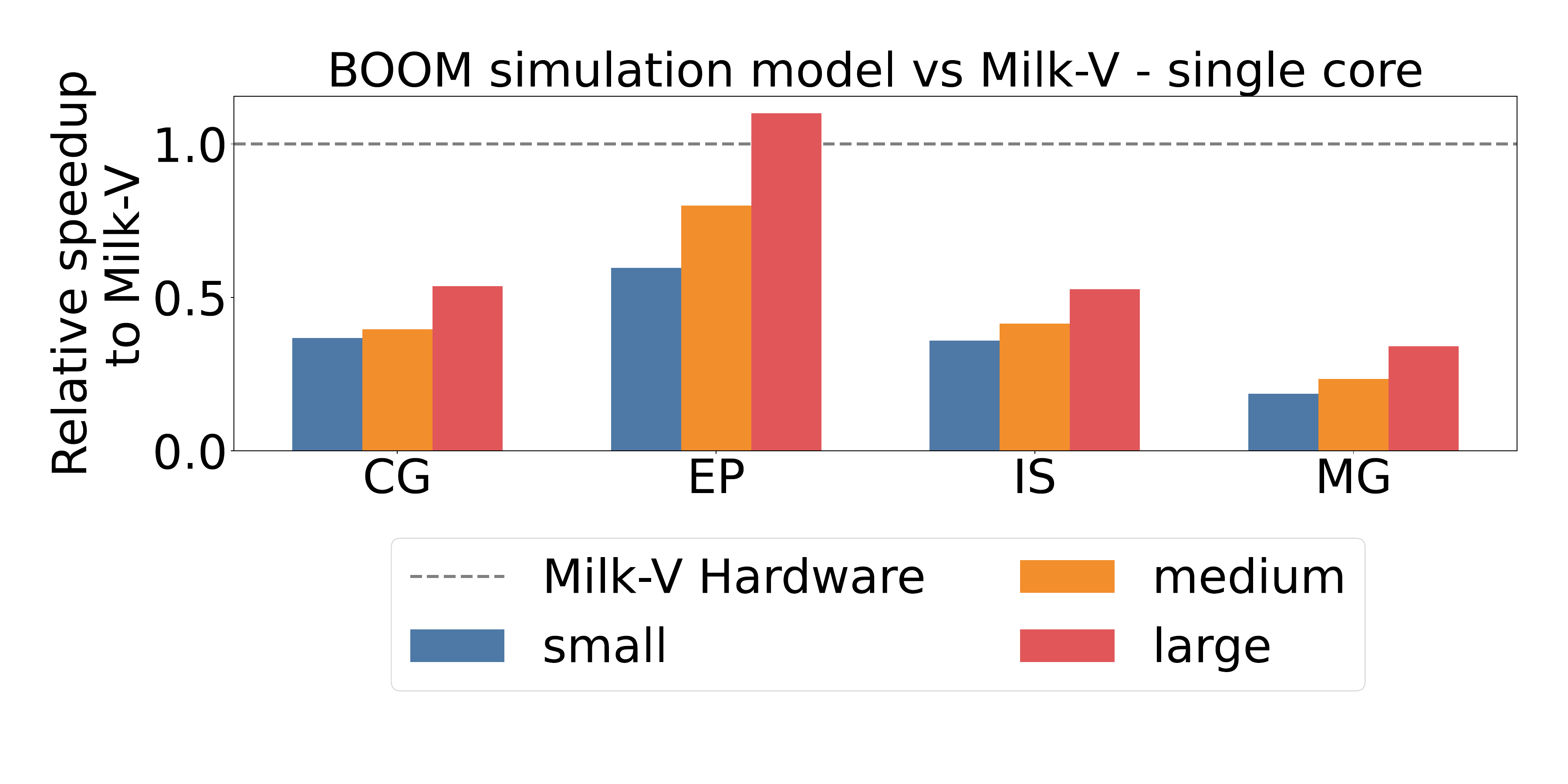}
  }
  \par\smallskip
  \subcaptionbox{MILK-V simulation model%
    \label{results:boom_npb:multi}}{%
    \includegraphics[width=0.5\textwidth]{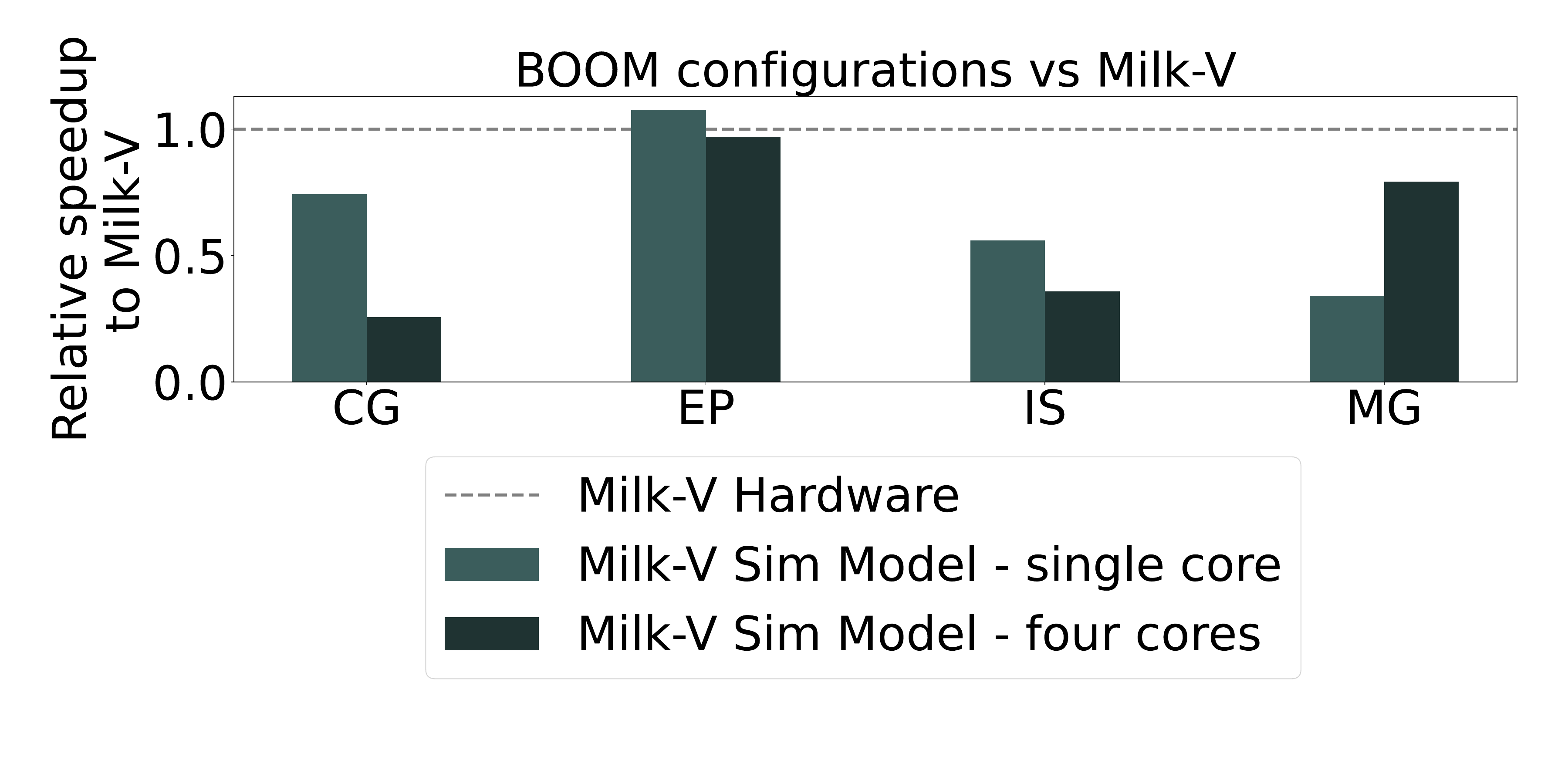}
  }

  \caption{Relative speedup of NAS Parallel Benchmarks on FireSim (BOOM cores) compared to MILK-V hardware (target value = 1.0).}
  \label{results:boom_npb}
\end{figure}

Figure~\ref{results:boom_npb:init} shows the relative speedup of each configuration compared to MILK-V. Analysis of the single-core EP benchmark (compute-bound) shows that the large BOOM configuration delivers performance close to MILK-V. In contrast, the other benchmarks are noticeably slower on the emulated hardware, primarily due to differences in cache sizes, memory latency, and bandwidth. Benchmarks such as CG, IS, and MG are all sensitive to changes in memory characteristics, whether in cache capacity, access latency, and overall bandwidth. 

Based on these findings, we selected the large BOOM configuration as our baseline and adjusted its cache parameters to match the MILK-V hardware specifications. Figure~\ref{results:boom_npb:multi} presents the results for single and four core runs, with the tuned configuration.

For the single-core case (Figure~\ref{results:boom_npb:multi}), increasing the L1 cache (from 32 KB to 64 KB) improved CG benchmark performance compared to the initial large BOOM configuration, reducing runtime by approximately 27.7 \%., and thus bringing it closer to the performance observed on real hardware. However, this change had little effect on the IS and MG benchmarks, which are limited by memory bandwidth. Both benchmarks still exhibited a substantial performance gap between the simulation model and the MILK-V hardware. Reducing this gap would require modeling DDR4 memory to match the actual MILK-V configuration, as the current simulation uses DDR3 and improving core (larger ld/st queue, larger reorder buffer size \emph{etc}.) as well as improving memory subsystem's capability (higher cache MSHRs, larger queue for DRAM \emph{etc}.).

For multi core runs, we observed an even larger performance gap for CG and IS. Benchmarks sensitive to cache latency and memory bandwidth showed poor strong scaling in the simulation compared to MILK-V, where scaling was notably better. An exception was the MG benchmark, which scaled poorly on MILK-V (the observed runtime on single and four cores was identical). We attribute this difference to the inability of BOOM, in simulation, to saturate the memory subsystem with a single core (allowing it to benefit from additional cores) whereas MILK-V already limited by memory subsystem performance in single-core runs, showing no performance gain when more cores are used. Consequently, the MG benchmark on four cores exhibited a smaller performance gap.

Finally, the EP benchmark demonstrated near performance parity between simulation and hardware for single core and four core runs. This confirms the compute capabilities of the large BOOM configuration are very close to those of the MILK-V hardware.



\subsection{UME}

\begin{figure}[htbp]
  \centering
  \includegraphics[width=\columnwidth]{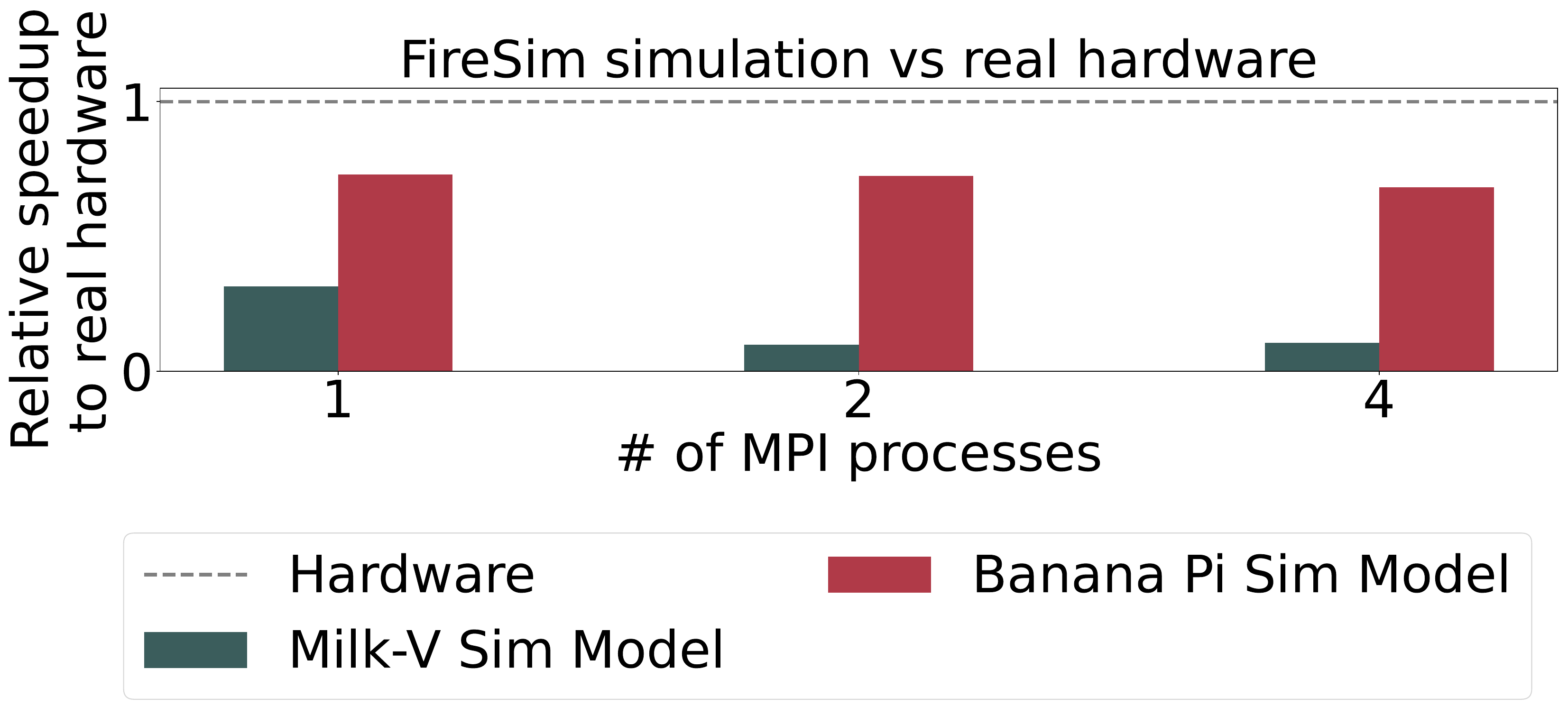}
  \caption{Relative speedup of UME on FireSim compared to actual hardware. Here Rocket-based Banana Pi Sim model is compared with Banana Pi hardware and BOOM-based MILK-V model is compared with MILK-V hardware.}
  \label{results:ume}
\end{figure}

Figure~\ref{results:ume} compares the relative speedup of the Rocket core-based simulation model (\texttt{Banana Pi Sim Model}) with the Banana Pi hardware and the BOOM-based simulation model (\texttt{MILK-V Sim Model}) with the MILK-V hardware while running UME. We add the times taken by the \emph{original kernel, inverted kernel, and face area calculation kernel} to compute the total runtimes for UME. The runtimes on Banana Pi are $0.73$, $0.4$, $0.21$ seconds for 1, 2, and 4 MPI processes while the runtimes of corresponding FireSim simulations are $1$, $0.56$, and $0.31$, closely matching the Banana Pi performance. On the MILK-V, the runtimes are $0.15$, $0.03$, $0.016$ seconds for 1, 2, and 4 ranks and while the runtimes of the corresponding FireSim simulation are $0.49$, $0.28$, and $0.15$ seconds, respectively. 
As expected, we observe runtime scaling with MPI ranks. Among the four tests (across simulations and hardware), we observe that the MILK-V significantly outperforms its corresponding FireSim simulation in terms of runtimes. We also observe that the Banana Pi performs better compared to the modified Rocket cores used to model it (\texttt{Banana Pi Sim Model}) which is expected since Banana Pi has 8-stage pipeline with dual-issues while the Rocket core has a 5-stage, single-issue pipeline.

\subsection{Lammps}

\begin{figure}[tbp]
  \centering
  \includegraphics[width=\columnwidth]{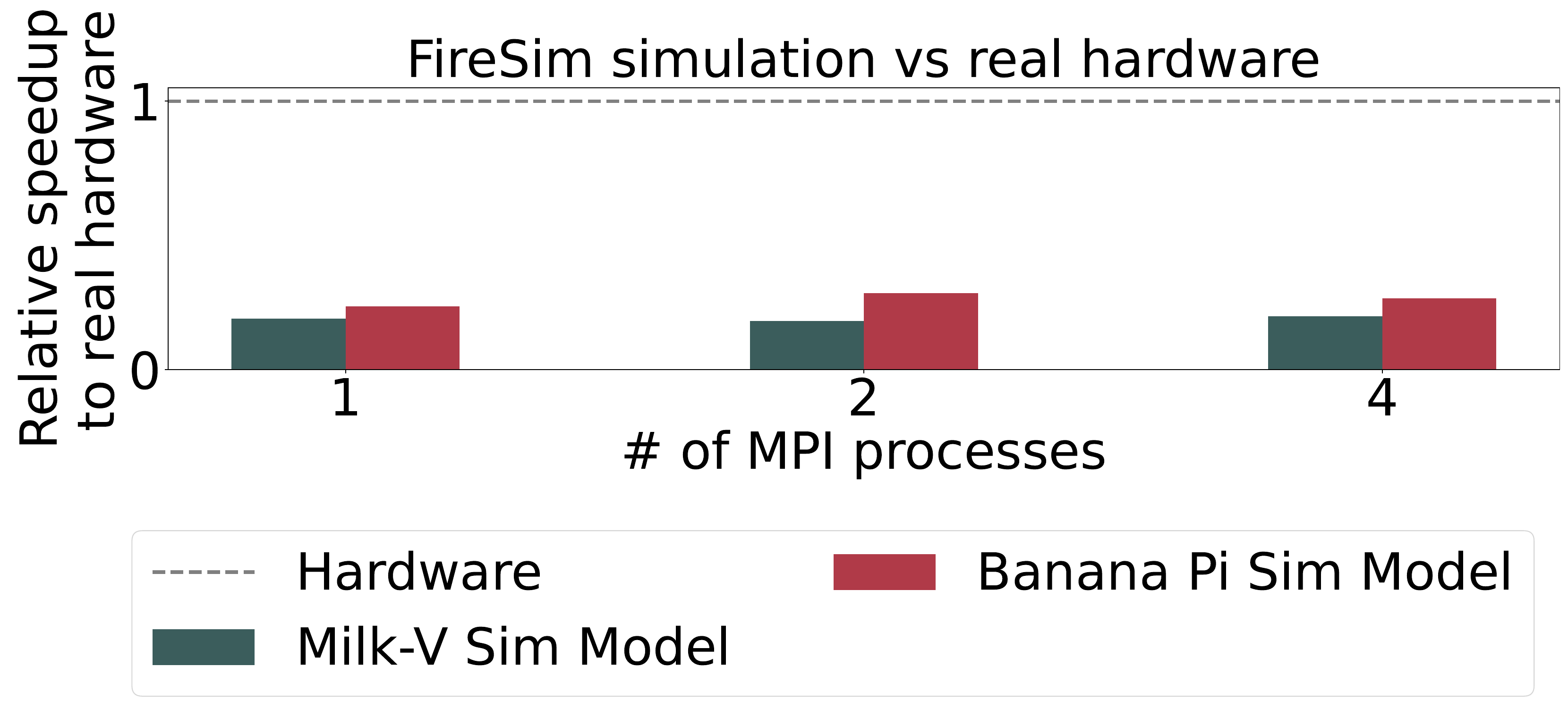}
  \caption{Relative speedup of Lammps-LJ on FireSim compared to actual hardware. Rocket-based \texttt{Banana Pi Sim Model} is compared with \texttt{Banana Pi hardware} and BOOM-based \texttt{MILK-V Sim Model} is compared with \texttt{MILK-V hardware}.}
  \label{results:lammps-lj}
\end{figure}

\begin{figure}[tbp]
  \centering
  \includegraphics[width=\columnwidth]{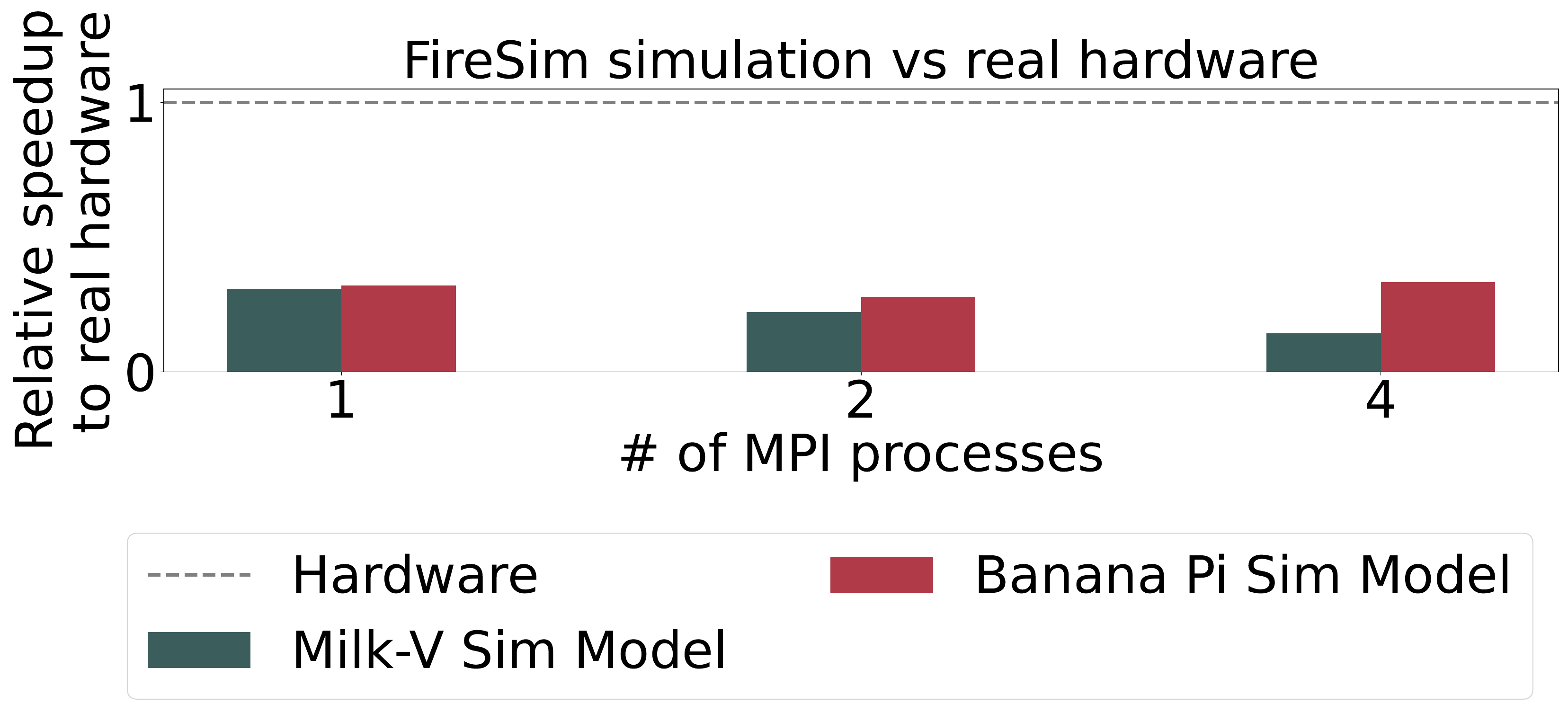}
  \caption{Relative speedup of Lammps-Chain on FireSim compared to actual hardware. Rocket-based \texttt{Banana Pi Sim Model} is compared with Banana Pi hardware and BOOM-based \texttt{MILK-V Sim Model} is compared with \texttt{MILK-V hardware}.}
  \label{results:lammps-chain}
\end{figure}

Figures~\ref{results:lammps-lj} and~\ref{results:lammps-chain} shows the relative speedup of Lennard-Jones (LJ) and Polymer Chain Lammps benchmarks (each has $32,000$ atoms and runs for $100$ timesteps) running on FireSim simulated platform compared to actual hardware. Similar to UME, we also observe speedup with the number of MPI processes. The LJ runtimes on Banana Pi are $13$, $8$, $4$ seconds for 1, 2, and 4 MPI processes while the runtimes of the corresponding FireSim simulation are $55$, $28$, and $15$, not closely matching Banana Pi. On MILK-V, the runtimes are $4$, $2$, $1$ seconds for 1, 2, and 4 ranks and $21$, $11$, and $5$ seconds for the FireSim simulation, indicating a large performance gap between MILK-V hardware and the corresponding FireSim model.

Similarly, the \emph{Chain} runtimes on Banana Pi are $9$, $5$, $4$ seconds for 1, 2, and 4 MPI processes while the runtimes of corresponding FireSim simulation are $28$, $18$, and $12$ seconds, highly deviating from Banana Pi. On MILK-V, the runtimes are $4$, $2$, $1$ seconds for 1, 2, and 4 ranks and $13$, $9$, and $7$ seconds for the FireSim simulation. Once again, there is a large performance gap between MILK-V hardware and FireSim simulation although good MPI performance scaling can be observed in all hardware configurations.

\section{Discussion}
Our results show that while the tuned FireSim models capture many performance characteristics of the Banana Pi and MILK-V hardware, achieving performance parity across all workloads remains challenging. Several factors contribute to this difficulty. First, the lack of complete microarchitectural specifications from hardware vendors prevents accurate replication of key parameters. Important details such as cache latencies, TLB sizes, load/store queue depths, and the number of cache banks are not publicly available. Second, some components cannot be modeled precisely within the current FireSim framework without significant engineering effort. For example, accurately modeling DDR4 would require a custom memory model, as FireSim currently supports only DDR3. Similarly, the Banana Pi employs a dual-issue, eight-stage pipeline, whereas the Rocket core model in FireSim is a single-issue, five-stage pipeline; replicating the former would demand substantial design modifications.

Software-based microbenchmarking proved valuable for guiding the tuning process. The MicroBench suite results identified specific areas where the simulation model diverged from the real hardware, such as control-flow execution, cache behavior, and memory bandwidth. These results informed targeted adjustments to cache sizes, bus widths, and other parameters. However, microbenchmark interpretation is not always straightforward. Many kernels stress multiple subsystems simultaneously. In such cases, deciding which parameters to modify for improved fidelity is inherently ambiguous.

Overall, our study highlights the strengths and limitations of FireSim for modeling commercially available RISC-V hardware. While the framework provides sufficient flexibility to approximate system-level performance trends and to explore architectural trade-offs, achieving cycle-level accuracy requires a combination of detailed hardware disclosure and extensive simulator enhancements. We provide raw runtime data on GitHub\footnote{\url{https://github.com/diehlpkpapers/risc-v-compare}}.

\section{Conclusion}
Understanding the necessary features required to bridge the gap between the performance of RISC-V hardware and FireSim simulations was the primary goal of this paper. To accomplish this, we utilized the single-board computer Banana Pi and the desktop-grade MILK-V Pioneer. We encountered the following challenges: \textit{1)} from the hardware perspective, not all details were provided by the manufacturer, \emph{e.g.}\ number of memory channels, which were required for the FireSim configuration and \textit{2)} not all features, such as a DDR4 memory model, are currently implemented in FireSim. These challenges made it difficult to match the exact hardware configurations in the simulation models. For the Banana Pi, the runtime on the hardware and FireSim match for some of microbenchmarks and NAS, while the performance of UME and LAMMPS diverged due to these limitations. For MILK-V, hardware and software simulation performance diverged due to the mismatch between DDR4 and DDR3. These results provided crucial insights into the current limitations of FireSim and additional micro-architecture details from the hardware vendors are needed to enable matching configurations. Furthermore, simulations on FireSim are time consuming, and we limited the run time of applications to a few minutes to ensure the simulations were able to complete in a few hours.  

Future work will focus on addressing the challenges involved in aligning performance between hardware and simulation. One key advantage of FireSim is its ability to simulate multiple nodes, enabling the execution of distributed runs. In future studies, simulations up to eight nodes can be performed in the available BxE environment, and further scaling studies can be performed using FPGAs available through AWS. 

\begin{acks}
We thank the Center of Computation \& Technology at Louisiana State University for providing us access to the MILK-V and Banana Pi hardware. This work was supported by the U.S. Department of Energy through the Los Alamos National Laboratory. Los Alamos National Laboratory is operated by Triad National Security, LLC, for the National Nuclear Security Administration of U.S. Department of Energy (Contract No. 89233218CNA000001). It has been approved for public release with LA-UR-25-27825.
\end{acks}


\printbibliography

\end{document}